\documentclass[showpacs,twocolumn,amsmath,amssymb,fixfloats]{revtex4}
\usepackage[english]{babel}
\usepackage{amsmath}
\usepackage{amssymb}
\usepackage{graphicx}

\begin{document}

\title{Classical dynamics of the optomechanical modes of a Bose-Einstein
condensate in a ring cavity}

\author{W. Chen}
\author{D. S. Goldbaum}
\author{M. Bhattacharya}
\author{P. Meystre}

\affiliation{B2 Institute, Department of Physics and College of Optical Sciences,\\
The University of Arizona, Tucson, AZ 85721 }

\begin{abstract}
We consider a cavity optomechanical system consisting of a Bose-Einstein condensate (BEC) interacting with two counterpropagating traveling-wave modes in an optical ring cavity. In contrast to the more familiar case where the condensate is driven by the standing-wave field of a high-$Q$ Fabry-P{\'e}rot cavity we find that both symmetric and antisymmetric collective density side modes of the BEC are mechanically excited by the light field. In the semiclassical, mean-field limit where the light field and the zero-momentum mode of the condensate are treated classically the system is found to exhibit a rich multistable behavior, including the appearance of isolated branches of solutions (isolas). We also present examples of the dynamics of the system as input parameters such as the frequency of the driving lasers are varied.
\end{abstract}
\maketitle

\section{\label{introduction} Introduction}

The optomechanical cooling of mechanical oscillators has witnessed considerable progress in the last few years, leading to the expectation that a large class of such oscillators will soon be cooled to near their quantum-mechanical ground state of vibration. In addition to the {\em top-down} approach where cooling  proceeds by mounting macroscopic oscillators as moving mirrors in an optical resonator \cite{vahala2008, mechq} --- often but not always the end-mirror of a Fabry-P{\'e}rot interferometer, there has also been increased interest in considering {\em bottom-up} situations. In that case the mechanical oscillator consists of a momentum side mode of an ultracold atomic system trapped inside a high-$Q$ optical cavity with fixed mirrors. The trapped atoms can be either a thermal sample \cite{murch2008}, a quantum-degenerate Bose-Einstein condensate (BEC) \cite{esslinger2008, esslinger2009}, or even a quantum-degenerate gas of fermions \cite{rina2010}. In that bottom-up situation the mechanical oscillator(s) consist of \textit{collective} momentum modes of the trapped gas, excited via photon recoil \cite{nagy2009,liu2009,keye2009,dan2009,aranya2009}. Specifically, in the case of a condensate the intracavity standing-wave field couples the macroscopically occupied zero-momentum component of the BEC to a symmetric superposition of the states with center-of-mass momentum  $\pm 2\hbar k$ via virtual electric dipole transitions.

It is known that when considering the mechanical effects of light on atoms by quantized light fields, there are situations where a standing wave does not lead to the same diffraction pattern as a superposition of two counterpropagating running waves of equal frequencies \cite{shore1991, dominic2005, mekhov2009}. This is because in contrast to a standing wave, running waves in principle permit one to extract ``which way'' information about the matter-wave diffraction process. This begs the question whether a description of the intracavity light field in terms of a standing wave is always appropriate to describe the optomechanical effects of feeble light fields on ultracold atoms. As a first step toward answering this question, this paper addresses the somewhat simpler question of understanding the difference between {\em classical} standing wave and counterpropagating light fields, that is, the difference in optomechanical properties of condensates trapped in, say, a Fabry-P{\'e}rot and a ring cavity. One main consequence of the presence of two counterpropagating running waves is that in addition to a symmetric ``cosine'' momentum side mode, it is now possible to excite an out-of-phase ``sine'' mode as well. In the optomechanics analogy, this indicates that two coupled ``condensate mirrors'' of equal oscillation frequencies but in general different masses are driven by the intracavity field. This can result in complex multistable behaviors that we analyze in some detail in the following sections.

The dynamical interaction between BECs and counterpropagating light fields has a long history. The cooperative scattering of light and atoms in ultracold atomic systems, including experimental \cite{Inouye1999,Schneble2003,Schneble2004,Yoshikawa2004} and theoretical \cite{Moore1999,Piovella2001,Robb2005,Zobay2005,Uys2007} studies of superradiance in BECs and coherent atom recoil lasing (CARL) \cite{Bonifacio1994,Bonifacio21994,Bonifacio1995,Bonifacio1997,Lippi1996,Hemmer1996,Kruse2003,Cube2004} has been the subject of a number of studies, see Ref.~\cite{Uys2008} for a brief review. The work most closely related to our analysis is Ref.~\cite{Elsasser2004}, which considers likewise the collective dynamics of atoms in a ring resonator, but in the somewhat simplified situation where the amplitude of one of the counterpropagating fields inside the resonator is fixed.

The remainder of the paper is organized as follows. Section II describes our model of a BEC interacting with two counterpropagating fields in a ring cavity. It introduces a description of the condensate that clearly illustrates an optomechanical analogy involving two mirrors driven by the intracavity field. In the semiclassical limit where the light fields are treated classically, the dynamics of these two ``mirrors'' can be understood in terms of an interference effect between the propagating fields. We further discuss conditions under which one of the mirrors can be trapped into a dark state. Section III discusses the steady-state properties of the system, and comments on the appearance of isolated branches of solutions, or ``isolas''~\cite{kaplan1985,pierre1986}. We then turn in Section IV to the dynamics of the system, considering in particular the response of the condensate to a sweep of the frequency of
the external pump lasers. Finally, Section V is a summary and outlook.

\section{\label{model} Model}

We consider a BEC of $N$ two-level bosonic atoms with transition frequency $\omega_a$ trapped inside one arm of a high-$Q$ optical ring cavity supporting two counterpropagating traveling-wave fields of frequency $\omega$ and driven by two coherent pump fields of identical frequency $\omega_p$ and complex amplitudes $\eta_1$ and $\eta_2$. The light fields are taken to be far detuned from the atomic transition so that the excited electronic state of the atoms can be adiabatically eliminated. We further assume that the atom trap is sufficiently soft that the condensate can be assumed to be homogeneous along the axis of the resonator. For simplicity we neglect transverse effects and describe both the condensate and the light fields as plane waves, thereby reducing the description of the problem to one dimension. Neglecting atom-atom collisions and in a frame rotating at the pump frequency $\omega_p$ the model Hamiltonian for this system is then
\begin{widetext}
\begin{eqnarray}
\hat H = -\sum_{i=1,2} \left[ \hbar \Delta \, \hat a_i^\dagger \hat a_i
+ i \hbar (\eta_i^* \hat a_i- \eta_i \hat a_i^\dagger) \right]
+ \int dx \, \hat \psi^\dagger(x) \left[ - \frac{\hbar^2}{2 m} \frac{d^2}{dx^2}
+ \hbar U_0 \left( \hat a_1^\dagger \hat a_1 + \hat a_2^\dagger \hat a_2
+ \hat a_2^\dagger \hat a_1 e^{2 i k x} + \hat a_1^\dagger \hat a_2 e^{-2 i k x} \right) \right]
\hat \psi(x), \nonumber \\
\label{h0}
\end{eqnarray}
\end{widetext}
where $\hat \psi(x)$ and $\hat a_i$ are the bosonic annihilation operators for the atomic field and the counterpropagating cavity modes $i=1,2$, respectively, $k=\omega/c$, $\Delta=\omega_p - \omega$ is the pump-cavity detuning, $m$ is the atomic mass, $U_0=g_0^2/\Delta_a$ is the familiar off-resonant atom-photon interaction strength, with $g_0$ the vacuum Rabi frequency and $\Delta_a=\omega_p-\omega_a$ the atom-pump detuning.

\begin{figure}
\includegraphics[width=7cm]{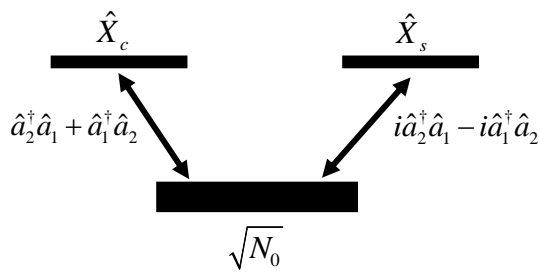}
\caption{Excitation of symmetric $\hat X_c$ and antisymmetric $\hat X_s$ collective density side modes.
The excitation of these modes is due to the interaction between a classically treated zero-momentum mode
$\hat c_0 \rightarrow \sqrt{N_0}$, and the superposition of two travelling wave modes $\hat a_1$ and $\hat a_2$. }
\label{sch}
\end{figure}

The BEC is initially prepared in a macroscopically occupied zero-momentum state from which the atoms are then scattered into higher momentum side modes $\pm 2 \ell \hbar k$ by the intracavity optical field, where $\ell$ is an integer. For moderate fields the condensate dynamics can be restricted to the zero-momentum mode and the first two side modes, $\ell = 1$ \cite{esslinger2008,nagy2009}, and the atomic field operator can be expanded simply in terms of these modes. Instead of a plane-wave expansion, we find it convenient to use a sine and cosine basis, so that
\begin{eqnarray}
\hat \psi(x)= \frac{1}{\sqrt{L}} \hat c_0+ \sqrt{\frac{2}{L}} \, \hat c_c \cos(2kx)
+ \sqrt{\frac{2}{L}} \, \hat c_s \sin(2kx),
\label{psi}
\end{eqnarray}
where $L$ is the cavity length and $\hat c_0$, $\hat c_c$ and $\hat c_s$ are the bosonic annihilation operators of the zero-momentum mode, the cosine mode, and the sine mode respectively. They satisfy the usual bosonic commutation relations $[c_i,c_j^{\dagger}]=\delta_{i,j}$. With this expansion of the Schr{\" o}dinger field the Hamiltonian \eqref{h0} reduces to
\begin{eqnarray}
\hat H&=& \hbar\left(-\Delta+ U_0 N \right) \sum \hat a_i^\dagger \hat a_i
-\sum i \hbar (\eta_i^* \hat a_i- \eta_i \hat a_i^\dagger)   \nonumber \\
&+& 4 \hbar \omega_r (\hat c_c^\dagger c_c+\hat c_s^\dagger c_s)
+\frac{\hbar \, U_0}{\sqrt{2}} (\hat a_2^\dagger \hat a_1+\hat a_1^\dagger \hat a_2)
(\hat c_c^\dagger \hat c_0+\hat c_0^\dagger \hat c_c)   \nonumber \\
&+&\frac{i\hbar \, U_0}{\sqrt{2}} ( \, \hat a_2^\dagger \hat a_1 - \, \hat a_1^\dagger \hat a_2)
(\hat c_s^\dagger \hat c_0+\hat c_0^\dagger \hat c_s ),
\label{heff}
\end{eqnarray}
where $\omega_r=\hbar k^2/2 m$ is the atomic recoil frequency. This Hamiltonian can be further simplified since for weak optical fields and large $N$ the depletion of the initial condensate remains weak. We then treat the zero-momentum mode classically, $\hat c_0 \rightarrow \sqrt{N}$ . We further introduce the `position' operators
\begin{equation}
\hat X_j \equiv (\hat c_j^\dagger+\hat c_j)/\sqrt{2}
\end{equation}
and the corresponding `momentum' operators
\begin{equation}
\hat P_j \equiv i (\hat c_j^\dagger-\hat c_j)/\sqrt{2},
\end{equation}
where $j=\{c,s\}$, resulting in the Hamiltonian \eqref{heff} being mapped to the optomechanical Hamiltonian
\begin{eqnarray}
\hat H&=& \hbar\left(- \Delta+  U_0 N \right) \sum \hat a_i^\dagger \hat a_i
-\sum i \hbar (\eta_i^* \hat a_i- \eta_i \hat a_i^\dagger)   \nonumber \\
&+& 2 \hbar \omega_r (\hat X_c^2+\hat P_c^2+\hat X_s^2+\hat P_s^2)  \nonumber  \\
&+& \hbar \, U_0 \sqrt{N} \hat X_c (\hat a_2^\dagger \hat a_1+\hat a_1^\dagger \hat a_2) \nonumber   \\
&+& i\hbar \, U_0 \sqrt{N} \hat X_s ( \hat a_2^\dagger \hat a_1- \hat a_1^\dagger \hat a_2).
\label{heff2}
\end{eqnarray}
In this form, the Hamiltonian provides a simple physical picture: the sine and cosine momentum side modes of the condensate behave as a pair of mirrors driven by the interference between the two counter-propagating intracavity light fields. As such, they can be thought of as a ``bottom-up'' realization of coupled mirrors driven by the radiation pressure. We return to this point shortly, but first derive the coupled equations of motion of the light-condensate mirrors system.

In practice the sine and cosine momentum modes are coupled to other momentum side modes as a result of the presence of a trapping potential that is otherwise ignored in our discussion. This results in a damping of the population of these modes, and associated noise operators. The resulting quantum Langevin equations obtained from the Hamiltonian~\eqref{heff2} are therefore
\begin{eqnarray}
\dot{\hat{X_c}}&=& 4 \omega_r \hat P_c - \gamma \hat X_c+\hat{f}_{xc},   \nonumber \\
\dot{\hat{P_c}}&=& -4 \omega_r \hat X_c - U_0 \sqrt{N} (\hat a_2^\dagger \hat a_1+\hat a_1^\dagger \hat a_2)
- \gamma \hat P_c +\hat{f}_{pc},   \nonumber \\
\dot{\hat{X_s}}&=& 4 \omega_r \hat P_s - \gamma \hat X_s+\hat{f}_{xs},  \nonumber  \\
\dot{\hat{P_s}}&=& -4 \omega_r \hat X_s - iU_0 \sqrt{N} ( \, \hat a_2^\dagger \hat a_1-  \, \hat a_1^\dagger \hat a_2)
- \gamma \hat P_s +\hat{f}_{ps}  \nonumber   \\
i \dot{\hat{a}}_1 &=& U_0 \sqrt{N} \hat a_2 (\hat X_c - i \hat X_s)-(\widetilde{\Delta}+ i \kappa) \hat a_1+ i \eta_1+\sqrt{2\kappa}\hat{a}_{1}^{\mathrm{in}},  \nonumber \\
i \dot{\hat{a}}_2 &=& U_0 \sqrt{N} \hat a_1 (\hat X_c + i \hat X_s)-(\widetilde{\Delta}+ i \kappa) \hat a_2+ i \eta_2+\sqrt{2\kappa}\hat{a}_{2}^{\mathrm{in}},\nonumber \\
\label{lang}
\end{eqnarray}
where $\kappa$ is the linewidth of the ring cavity, $\widetilde{\Delta}=\Delta-U_{0}N$ is the Stark-shifted cavity-pump detuning,
and the side-mode damping rate $\gamma$ affects both position and momentum, as discussed in Ref.~\cite{keye2009}.
All noise operators, $\hat{f}_{xc}$, $\hat{a}_1^{\rm in}$, etc. are assumed to have zero mean.

In the remainder of this paper we consider the simple limit where all fields, optical and matter-wave, are treated classically. A full quantum description of the problem will be considered in future work. Taking the expectation values of Eqs.~(\ref{lang}) and factorizing all operator products, for example $\langle {\hat a}_2 {\hat X}_c\rangle \rightarrow \langle {\hat a}_2 \rangle \langle {\hat X}_c \rangle$, $\langle {\hat a}^\dagger_2 {\hat a}_1 \rangle \rightarrow \langle {\hat a}^\dagger_2 \rangle \langle {\hat a}_1 \rangle$, etc, yields the classical equations of motion
\begin{eqnarray}
\label{eqxc}
\ddot{X_c}&=& -2 \, \gamma \, \dot{X_c}- (16 \, \omega_r^2+ \gamma^2) \, X_c \nonumber \\
\label{eqxc}
&& - 4 \, \omega_r \, U_0 \sqrt{N} \, (\alpha_2^* \, \alpha_1+ \alpha_1^* \, \alpha_2), \\
\label{eqxs}
\ddot{X_s}&=& -2 \, \gamma \, \dot{X_s}- (16 \, \omega_r^2+ \gamma^2) \, X_s  \nonumber \\
&& - 4 \,i \, \omega_r \, U_0 \sqrt{N} \,(\alpha_2^* \, \alpha_1- \alpha_1^* \alpha_2),  \\
\label{eqa1}
i \dot{\alpha}_1 &=& U_0 \sqrt{N} \alpha_2 (X_c - i X_s)-(\widetilde{\Delta}+ i \kappa) \alpha_1+ i \eta_1,   \\
i \dot{\alpha}_2 &=& U_0 \sqrt{N} \alpha_1 (X_c + i X_s)-(\widetilde{\Delta}+ i \kappa) \alpha_2+ i \eta_2,
\label{eqa2}
\end{eqnarray}
where we have set $\langle \hat a_i \rangle \rightarrow \alpha_i$, $\langle {\hat X}_c \rangle \rightarrow X_c$, etc.

Equations \eqref{eqxc} and \eqref{eqxs} are particularly instructive in that they show that the sine and cosine side modes are driven by out-of-phase components of the interference pattern produced by the counterpropagating optical fields. In particular, we observe that if these two fields have equal phase and amplitude, then the sine mode is not excited: it becomes a dark state as a result of the destructive interference between the coherent recoil effects from the two counterpropagating fields. Similarly, the cosine mode can become a dark state for out-of-phase and equal amplitude intracavity fields, see Fig.~1. We show in the next section that these dark states are useful in understanding the steady-state properties of the system. Interestingly, in high-$Q$ Fabry-P{\'erot} resonators it is usually appropriate to describe the intracavity light field as a standing-wave, a cosine-wave if the origin is chosen as the center of the resonators. In that case, only the cosine momentum side mode of the condensate is excited, and the problem reduces to the familiar BEC optomechanical situation of Ref.~\cite{esslinger2008}.

We remark that even in Fabry-P{\'e}rot interferometers it is strictly speaking never completely appropriate to ignore the other mode, the sine mode of the optical field, unless the mirrors are perfectly reflecting (and infinitely heavy). This is especially so in the limit of feeble quantized fields, certainly the quantum fluctuations of the cosine mode are always present. It will be interesting in future work to determine the extent to which the two BEC `mirrors' under consideration here can develop quantum correlations between the two light fields, leading e.g. to $\langle \hat a_2^\dagger \hat a_1 \rangle \neq 0$, with similar behavior for the sine and cosine side modes. As such, this system may provide a sensitive test-bed to generate and study the onset of quantum correlations in this optically driven Bose condensate. These intriguing questions will be the topic of subsequent work. In the context of a BEC driven inside a Fabry-P{\'e}rot 
resonator, work has been done to treat quantum noise and correlations, see for instance \cite{nagy2009, Szirmai2009, Szirmai2010}. 
Concentrating for now on the classical properties of the system, we turn in the next section to a discussion of its steady-state properties.

\section{\label{Steady-state} Steady-state solutions and optical bistability}

Setting the time derivatives to zero in Eqs.~\eqref{eqxc}-\eqref{eqa2} yields after some algebra a fifth-order polynomial equation for $|\alpha_1|^2$. This equation is quite cumbersome and is not presented here. We proceed by numerically determining $|\alpha_1|^2$, retaining only the physically relevant real positive roots of the fifth-order equation. From these values one readily finds $|\alpha_2|^2$, $X_c$, $X_s$, $\alpha_1$ and $\alpha_2$ from
\begin{eqnarray}
\label{sta2}
\alpha_2 &=& \frac{i \eta_2}{\widetilde{\Delta}+ i \, \kappa+ C_1 \, |\alpha_1|^2 }, \\
\label{sta1}
\alpha_1 &=& \frac{i \eta_1}{\widetilde{\Delta}+ i \, \kappa+ C_1 \, |\alpha_2|^2 }, \\
\label{stxc}
X_c &=& C_2 (\alpha_2^* \, \alpha_1+ \alpha_1^* \, \alpha_2), \\
\label{stxs}
X_s &=&  i \, C_2 (\alpha_2^* \, \alpha_1- \alpha_1^* \, \alpha_2), \\ \nonumber
\end{eqnarray}
where
$$
C_1=8 \omega_r N U_0^2/(16 \omega_r^2+ \gamma^2)
$$
and
$$
C_2=-4 \omega_r U_0 \sqrt{N}/(16 \omega_r^2+ \gamma^2).
$$
The stability of the steady state solutions is determined by the standard Routh-Hurwitz criterion \cite{kauffmann1987}.

\begin{figure}
\includegraphics[width=7cm]{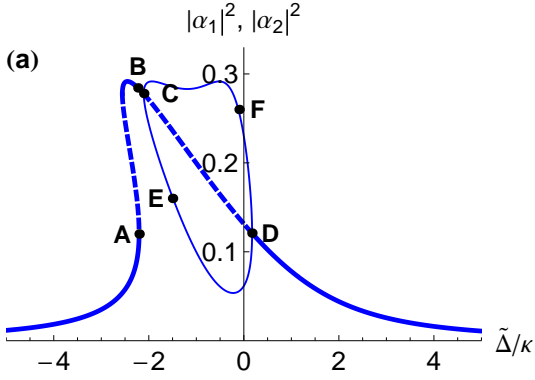}
\caption{ Mean intracavity intensity(unitless) of mode 1 or 2 as a function of the detuning $\widetilde{\Delta}$ for $L=100 \rm{nm}$, $\lambda_p=780 \rm{nm}$, $\kappa=2 \pi \times 1.3 \rm{MHz}$, $\omega_r=2 \pi \times 3.8 \rm{kHz}$, $U_0=2 \pi \times 3.1 \rm{kHz}$, $N=9000$, $\eta_1=\eta_2= 0.54 \kappa$, and $\gamma=0.001 \kappa$. The dashed segments are unstable. The plot for modes 1 and 2 are identical, except that when mode 1 is on the CFD branch then mode 2 is on the CED branch, and conversely, see text.}
\label{in1eq}
\end{figure}

\begin{figure}
\includegraphics[width=7cm]{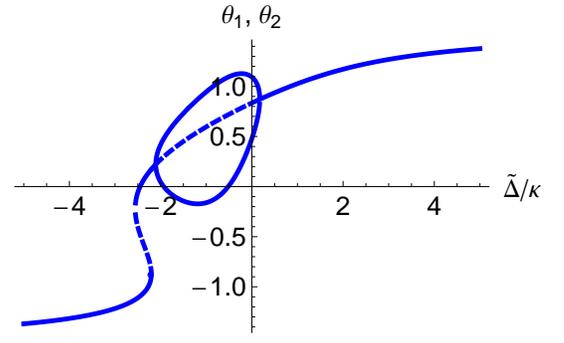}
\caption{ Phase of the intracavity fields as a function of the detuning $\widetilde{\Delta}$ for the same parameters as Fig.~2. The plots for modes 1 and 2 are identical, except that when mode 1 is on the upper semi-loop then mode 2 is on the lower semi-loop, and conversely, see text. The dashed segments are unstable.}
\label{phase1eq}
\end{figure}

\begin{figure}
\includegraphics[width=8cm]{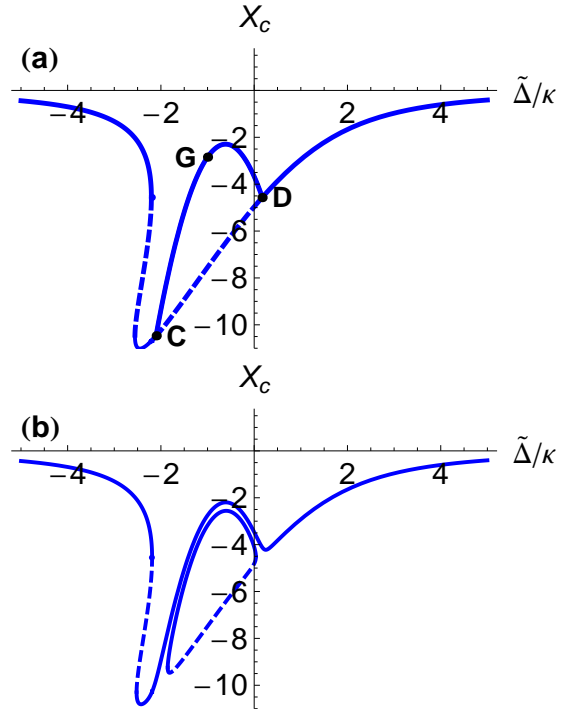}
\caption{ Position $X_c$(unitless) of the cosine mode as a function of the detuning $\widetilde{\Delta}$.(a) Equal pumping amplitudes $\eta_1=\eta_2$; (b) slightly imbalanced pumping $\eta_1= 0.53 \kappa$; $\eta_2= 0.54 \kappa$; other parameters as those in Fig.~\ref{in1eq}. The dashed segments are unstable.}
\label{xceq}
\end{figure}

\begin{figure}
\includegraphics[width=8cm]{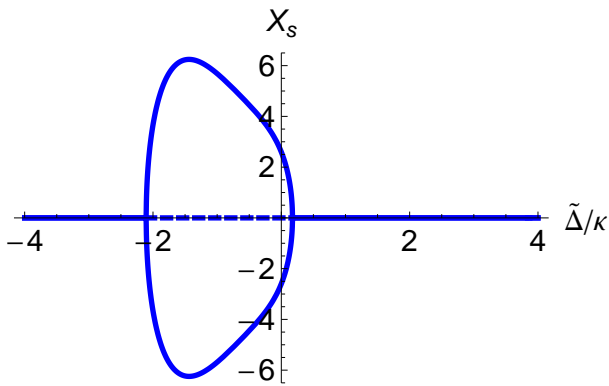}
\caption{ Position $X_s$(unitless) of the sine mode as a function of the detuning $\widetilde{\Delta}$ for equal pumping amplitudes. Parameters as in Fig.\ref{in1eq}. The dashed segments are unstable.}
\label{xseq}
\end{figure}

\begin{figure}
\includegraphics[width=7cm]{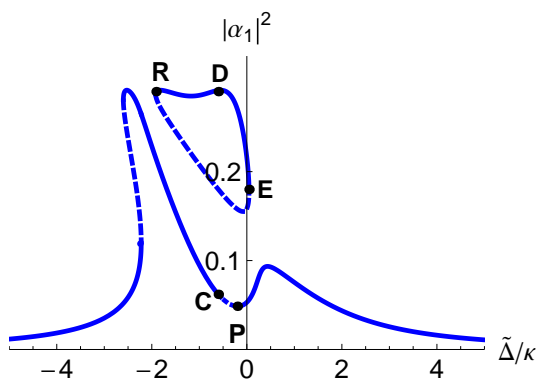}
\caption{ Mean intracavity intensity(unitless) of mode 1 as a function of detuning $\widetilde{\Delta}$. Parameters as those in Fig.\ref{in1eq} except that $\eta_1= 0.54 \kappa$ and $\eta_2= 0.55 \kappa$. The dashed segments are unstable. }
\label{in1neq}
\end{figure}

\begin{figure}
\includegraphics[width=7cm]{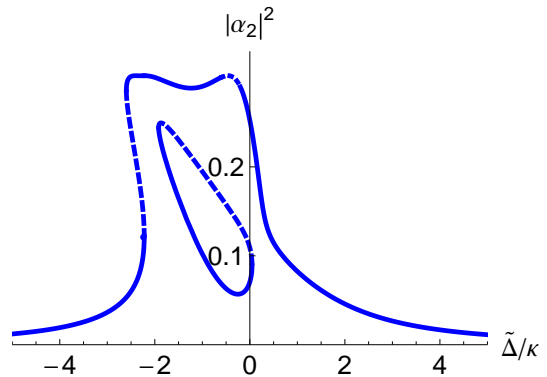}
\caption{ Mean intracavity intensity(unitless) of mode 2 as a function of detuning $\widetilde{\Delta}$. Same parameters as those in Fig.\ref{in1neq}. The dashed segments are unstable.}
\label{in2neq}
\end{figure}

\begin{figure}
\includegraphics[width=7cm]{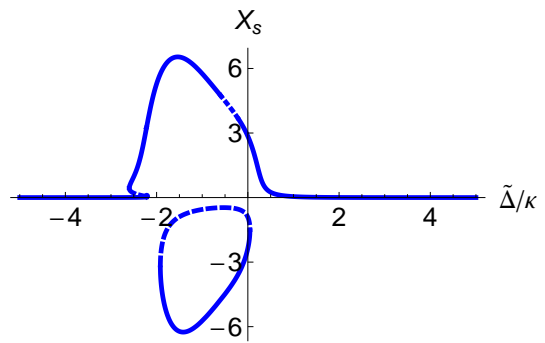}
\caption{ Position $X_s$(unitless) of the sine mode as a function of detuning $\widetilde{\Delta}$. Same parameters as those in Fig.\ref{in1neq}. The dashed segments are unstable. }
\label{xsneq}
\end{figure}

Figures \ref{in1eq} - \ref{xsneq} show typical results obtained for parameters similar to those used in previous experimental~\cite{esslinger2008} and theoretical~\cite{keye2009} work. For these specific parameters we obtain only three real roots for $|\alpha_1|^2$.

Consider first the case of real and equal pump amplitudes, $\eta_1=\eta_2$.
Figure~\ref{in1eq} shows the mean intracavity intensity of the counterpropagating optical fields 1 and 2 as a function of the effective detuning $\widetilde{\Delta}$. Starting from large negative detunings, the intracavity field intensities increase as the magnitude of the detuning is decreased, as expected, with both field intensities being equal. At point A the system makes a discontinuous jump to point B.
A similar behavior has been observed e.g. in the case of a BEC trapped in a Fabry-P{\'e}rot cavity \cite{esslinger2008}. However, a new feature of the ring cavity system appears at point C. Here, the solution described by the thick curve becomes unstable and the system undergoes a {\em spontaneous symmetry breaking}, with one of the intracavity intensities increasing and following the CFD curve, while the other decreases and follows the CED curve. Which of the two counterpropagating fields will follow which branch is completely random and determined by classical (or quantum) noise, as illustrated in the following section. This behavior is reminiscent of that found for instance in symmetrically pumped nonlinear interferometers \cite{kaplan1982, pierre1986}.

Figure~\ref{xceq}(a) shows the steady-state position $X_c$ of the cosine mode for the parameters of Fig.~\ref{in1eq}. Clearly, $X_c$ is also multistable, although this would appear not to be the case from the figure. This is because the branch CGD is actually degenerate, with the field intensities corresponding to both branches CED and CFD in Fig.~\ref{in1eq} yielding the same value of $X_c$ for a given detuning $\widetilde{\Delta}$. This might appear surprising, considering that these intensities are vastly different. The point is that the intensity only gives part of the story; we also need to consider the phase of the intracavity fields, see Fig.~3, which are just as important in determining the steady-state value of $X_c$ and $X_s$, as apparent from Eqs.~(\ref{stxc}) and (\ref{stxs}). More precisely, it is the interference between the forward and backward field amplitudes that drives $X_c$ and $X_s$. To the left of the spontaneous symmetry breaking point C in Fig.~\ref{in1eq} the phases of the two counterpropagating fields are equal, hence $\alpha_2^* \alpha_1 - \alpha_2 \alpha_1^* =0$, and the sine side mode is in a dark state, cf. Fig.~5. The situation is the same to the right of point D in Fig.~2. For detunings in the spontaneous symmetry breaking region, the upper intensity branch corresponds to the lower phase branch. For a fixed detuning, these combinations of phase and amplitudes of the fields result in equal values of the interference term $\alpha_2^* \alpha_1 + \alpha_2 \alpha_1^*$ for the two branches,  and hence the same value of $X_c$. In this region we also have that $\alpha_2^* \alpha_1 - \alpha_2 \alpha_1^* \neq 0$, resulting in non-zero values for $X_s$, see Fig.~\ref{xseq}. Note however that for that mode, the two branches lead to two values of $X_s$ exactly out of phase with each other, as expected from the form of Eq.(\ref{stxs}).

A small imbalance between the two (real) pump field amplitudes, $\eta_1 \neq \eta_2$ is sufficient to lift the degeneracy and remove the symmetry between the two counterpropagating modes, see Fig.~4(b). This leads to a qualitatively different steady-state behavior of the system, including the appearance of isolated domains of solutions, or isolas. A similar behavior has been previously predicted in other nonlinear optics contexts, including nonlinear ring resonators filled by a Kerr nonlinear medium \cite{kaplan1985,pierre1986}.

\begin{figure}
\includegraphics[width=8cm]{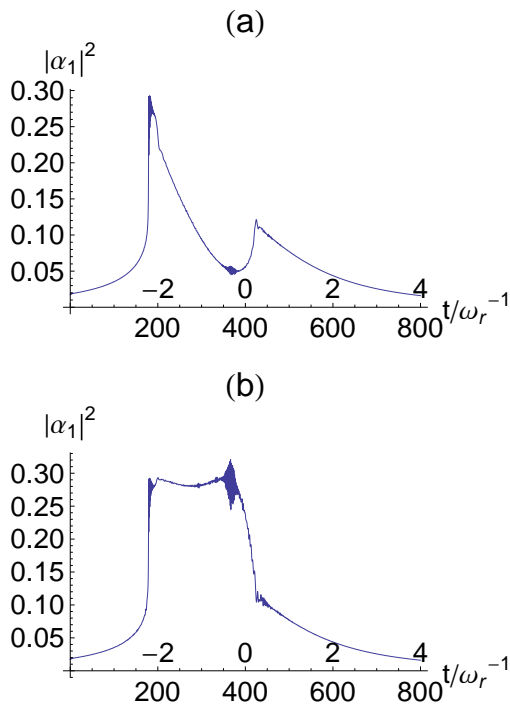}
\caption{Time evolution of the intensity(unitless) of mode 1 as the effective detuning
is varied at a rate of $2 \pi \times 0.3 \rm{MHz}$ from $-4 \kappa$ to $4 \kappa$. Due to random noise, the system evolves either along (a) the lower stable loop or (b) the upper stable loop, cf. Fig.\ref{in1eq}. The values given above the time axis are the instantaneous values of the detuning in units of $\kappa$. Other parameters as in Fig.\ref{in1eq}. }
\label{dya1sqeq}
\end{figure}

\begin{figure}
\includegraphics[width=8cm]{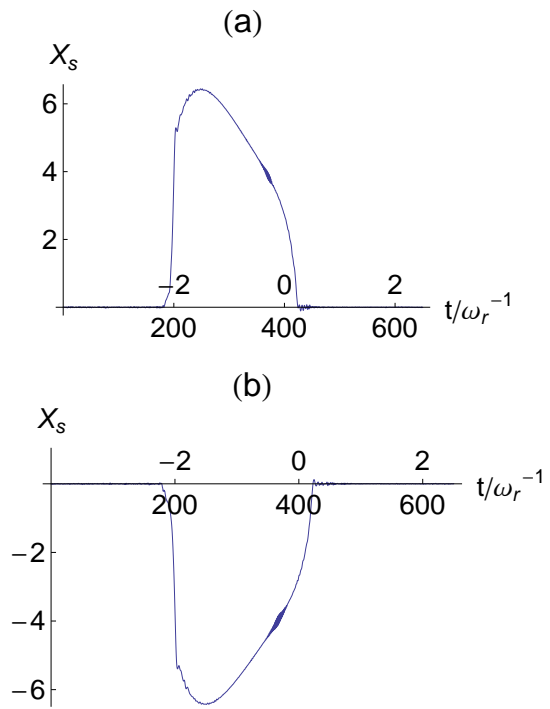}
\caption{Time evolution of $X_s$(unitless) as the effective detuning
is swept linearly at a rate of $2 \pi \times 0.3 \rm{MHz}$ from $-4 \kappa$ to $4 \kappa$. Classical noise can result in the system evolving either along (a) the upper or (b) the lower stable loop, see Fig.\ref{xseq}. The values given above the time axis are the instantaneous values of the detuning in units of $\kappa$. Other parameters as in Fig.\ref{in1eq}. }
\label{dyxseq}
\end{figure}

\begin{figure}
\includegraphics[width=9cm]{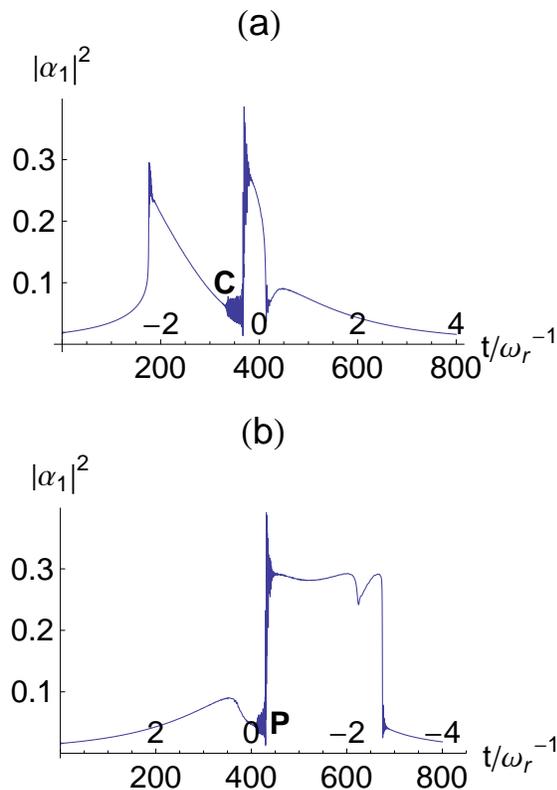}
\caption{ Time evolution of the intensity(unitless) of mode 1 under a linear sweep of the effective detuning at a rate of $2 \pi \times 0.3 \rm{MHz}$, (a) for a sweep from $-4 \kappa$ to $4 \kappa$; (b) for a sweep from $4 \kappa$ to $-4 \kappa$, cf. Fig.\ref{in1neq}. The values given above the time axis are the instantaneous values of the detuning in units of $\kappa$. Other parameters as in Fig.\ref{in1neq}.}
\label{dya1sqneq}
\end{figure}

\begin{figure}
\includegraphics[width=8cm]{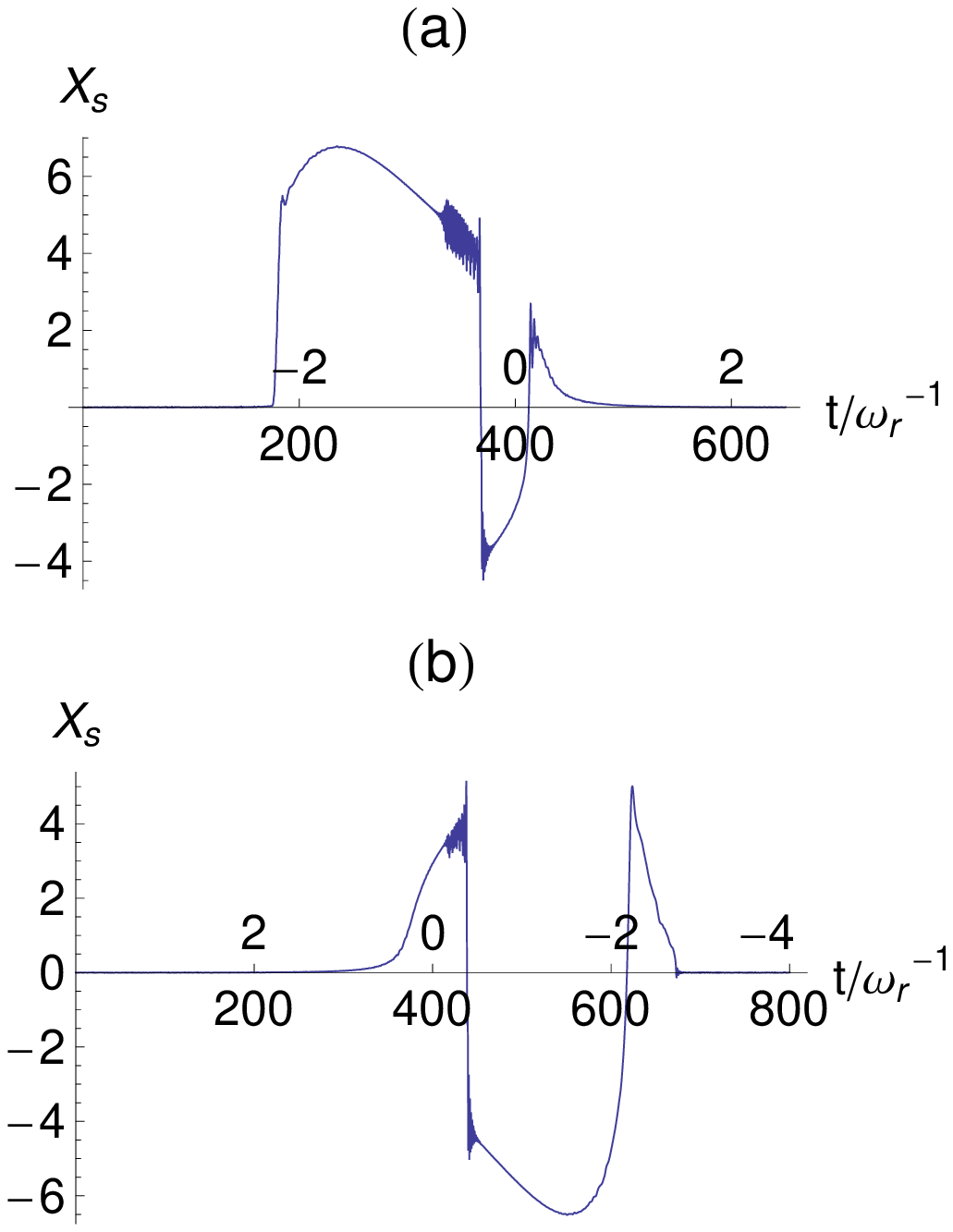}
\caption{ Time evolution of $X_s$(unitless) as the effective detuning is swept linearly at a rate of $2 \pi \times 0.3 \rm{MHz}$. (a)
from $-4 \kappa$ to $4 \kappa$; (b) from $4 \kappa$ to $-4 \kappa$, cf. Fig.\ref{xsneq}. The values given above the time axis are the instantaneous values of the detuning in units of $\kappa$. Other parameters as in Fig.\ref{in1neq}.}
\label{dyxsneq}
\end{figure}

Figure~\ref{in1neq} shows the mean intracavity intensity of mode 1 as a function of the detuning $\widetilde{\Delta}$ for imbalanced pumping, and Fig.~\ref{in2neq} the corresponding curves for mode 2. The effect of imbalanced pumping is to detach the loop previously associated with spontaneous symmetry breaking from the ``main'' bistability branch. The isolated branch, the isola, is accessible via an adiabatic sweep of the detuning from large negative values. When reaching point C in Fig.6 the main branch becomes unstable, and the field jumps toward point D on the isola. One could also reach the isola with a hard non-equilibrium excitation to a point within the basin of attraction of the steady-state branch RDE. That basin of attraction can be determined numerically by integrating the equations of motion (\ref{eqxc})-(\ref{eqa2}) backward in time from initial values arbitrarily close to that branch.

Fig.~\ref{xceq}(b) shows $X_c$ as a function of detuning, illustrating particularly clearly the lifting of the degeneracy associated with symmetric pumping and the onset of an isola, and Fig.~\ref{xsneq} the corresponding $X_s$.  Further results not discussed here show that when the difference between the pump field intensities are sufficiently large, both bistability and the isolas eventually disappear.

\section{\label{Dynamics} Dynamics}

We now turn to a discussion of selected aspects of the dynamics of the system as external parameters are varied in time. Here we concentrate on linear sweeps of the effective detuning from large negative values to large positive values.
Because the cavity field decay rate $\kappa$ is typically orders of magnitude larger than the characteristic frequency of mechanical motion $4 \omega_r$ and the decay rate $\gamma$ of the condensate, we can eliminate the optical field degrees of freedom adiabatically, and are left with a pair of coupled nonlinear second-order differential equations for $X_c$ and $X_s$ that can easily be solved numerically. The initial conditions for the oscillator variables are taken to be $X_{c}(0)=X_{s}(0)=\dot{X}_{c}(0)=\dot{X}_{s}(0)=0$ in all cases. We also add a small amount of classical noise $\delta N(t)$ to simulate atom number fluctuations, due e.g. to collisions or to thermal effects in the condensate. This noise term has the main effect of helping drive the system away from unstable branches in a reasonable time. In the simulations presented in this section $\delta N(t)$ is assumed to have a normal distribution with zero mean and the standard deviation equal to $5 \%$ of the total mean atom number.

We consider first the case of real pumps of equal amplitudes, $\eta_1=\eta_2$, and sweep the effective detuning linearly at a rate of $2 \pi \times 0.3 \rm{MHz}$, a rate that is in most cases slow compared to the characteristic time scale of the atomic motion, in order to guarantee adiabaticity, but still fast compared to the atom loss rate \cite{esslinger2008}. Figure~\ref{dya1sqeq} shows the time evolution of the intensity of mode 1 as the detuning $\widetilde{\Delta}$ is varied from $-4 \kappa$ to $4 \kappa$. Except for rapid transients that occur for a narrow range of detunings the intensity follows the steady-state values of Fig~.\ref{in1eq}. These rapid transients correspond to system parameters such that the system is almost unstable, with a particularly slow relaxation time to equilibrium. At the spontaneous symmetry breaking point C in Fig.2, random noise determines whether the intensity follows the lower half loop solution of Fig.~\ref{dya1sqeq}(a) or the upper half loop shown in Fig.\ref{dya1sqeq}(b). Figure~\ref{dyxseq} shows the corresponding $X_s(t)$.

We already mentioned that sweeping the effective detuning allows one to reach the isolas characteristic of imbalanced pumping. This is illustrated at point C of Fig.~\ref{dya1sqneq}(a), which shows the evolution of the intensity of mode 1 as the detuning is swept linearly from $-4 \kappa$ to $4 \kappa$. Likewise, point P on Fig.~\ref{dya1sqneq}(b) shows that transition when the detuning is swept down from $4 \kappa$ to $-4 \kappa$. For completeness, Fig.~\ref{dyxsneq} shows the corresponding time evolution of the position $X_s$ of the sine mode.

\section{Conclusions}
In this paper we have presented a classical analysis of the fundamental optomechanical modes of a Bose-Einstein condensate trapped inside a ring cavity. In contrast to the situation of a Fabry-P{\'e}rot cavity there are now two such modes, symmetric(cosine) and antisymmetric(sine) ones. These sine and cosine modes act as two coupled ``condensate mirrors'' of equal frequencies, their coupling resulting in a rich steady-state behavior, including for instance the appearance if isolas for an appropriate choice of parameters.

One important feature of these modes is that together with the original condensate they form a V-system, with the upper levels -- the sine and cosine modes -- driven by a two-photon process involving both counterpropagating light fields. One or the other of these modes can then become a dark state as a result of destructive interferences between the two counterpropagating fields. This is of course a classical effect: in the case of quantized optical and matter-wave fields quantum fluctuations will normally prevent these modes from becoming perfectly dark. It follows that measuring correlation functions of the optical field provides a direct means to probe the quantum properties of the matter-wave side modes. Future work will examine these aspects of ring-cavity optomechanics in detail, both when quantum fluctuations can be treated as small fluctuations about a classical mean, and in the limit of very feeble fields where this approach is no longer appropriate. We will also revisit the Fabry-P{\' e}rot case to include both the standing-wave mode that is normally included in the analysis, as well as the second frequency-degenerate but out-of-phase optical mode that needs to also be included and can become significantly excited for low mirror reflectivities. Finally, we will investigate in detail the damping of the matter-wave side modes due to the presence of a trap as well as the damping due to collisions. This will provide us with an effective $Q$-factor for the condensate mirrors and permit us to study in detail their heating and cooling mechanisms.

\section*{Acknowledgements}
We thank D. Meiser and K. Zhang for useful discussions. This work is supported in partby the US Office of Naval Research, by the National Science Foundation, and by the US Army Research Office.


\begin{thebibliography}
\expandafter\ifx\csname natexlab\endcsname\relax\def\natexlab#1{#1}\fi
\expandafter\ifx\csname bibnamefont\endcsname\relax
  \def\bibnamefont#1{#1}\fi
\expandafter\ifx\csname bibfnamefont\endcsname\relax
  \def\bibfnamefont#1{#1}\fi
\expandafter\ifx\csname citenamefont\endcsname\relax
  \def\citenamefont#1{#1}\fi
\expandafter\ifx\csname url\endcsname\relax
  \def\url#1{\texttt{#1}}\fi
\expandafter\ifx\csname urlprefix\endcsname\relax\def\urlprefix{URL }\fi
\providecommand{\bibinfo}[2]{#2}
\providecommand{\eprint}[2][]{\url{#2}}

\bibitem[{vah()}]{vahala2008}
\bibinfo{note}{T.J. Kippenberg and K. J. Vahala, Science
  \textbf{321}, 1171 (2008); F. Marquardt and S. Girvin, Physics,
  \textbf{2}, 40 (2009).}

\bibitem{mechq}
C. Hohberger-Metzger and K. Karrai, Nature {\bf 432}, 1002 (2004);
S. Gigan \textit{et. al}, Nature {\bf 444}, 67 (2006); O. Arcizet \textit{et. al} Nature {\bf 444}, 71 (2006); D. Klecker and D. Bouwmeester, Nature {\bf 444}, 75 (2006); A. Schliesser, P. Del'Haye, N. Nooshi, K. J. Vahala, and T. J. Kippenberg, Phys. Rev. Lett. {\bf 97}, 243905 (2006); M. Poggio, C. L. Degen, H. J. Mamin, and D. Rugar, Phys. Rev. Lett. {\bf 99}, 017201 (2007); I. Wilson-Rae, N. Nooshi, W. Zwerger, and T. J. Kippenberg, Phys. Rev. Lett. {\bf 99}, 093901 (2007); F. Marquardt, J. P. Chen, A. A. Clerk, and S. M. Girvin, Phys. Rev. Lett. {\bf 99}, 093902 (2007); C. Regal, J. Teufel, and K. Lehnert, Nature Physics {\bf 4}, 555 (2008); J. D. Thompson, B. M. Zwickl, A. M. Jayich, F. Marquardt, S. M. Girvin, and J. G. E. Harris, Nature {\bf 452}, 72 (2008).

\bibitem{murch2008}
K. W. Murch, K. L. Moore, S. Gupta, and D. M. Stamper-Kurn, Nature Physics {\bf 965}, 561 (2008).

\bibitem{esslinger2008}
F. Brennecke, S. Ritter, T. Donner and T. Esslinger, Science {\bf 322}, 235 (2008).

\bibitem{esslinger2009}
K. Baumann, C. Guerlin, F. Brennecke and T. Esslinger, arXiv:0912.3261 (2009).

\bibitem{rina2010}
R. Kanamoto and P. Meystre, Phys. Rev. Lett., in press

\bibitem{nagy2009}
D. Nagy, P. Domokos, A. Vukics, and H. Ritsch, Eur.Phys. J. D {\bf 55}, 659 (2009).

\bibitem{liu2009}
J. M. Zhang, F. C. Cui, D. L. Zhou, and W. M. Liu, Phys. Rev. A, {\bf 79}, 033401 (2009).

\bibitem{keye2009}
K. Zhang, W. Chen, M. Bhattacharya, and P. Meystre, Phys. Rev. A, {\bf 81}, 013802 (2010).

\bibitem{dan2009}
D. S. Goldbaum1, K. Zhang, and P. Meystre, arXiv:0911.3234 (2009).

\bibitem{aranya2009}
A. Bhattacharjee, Phys. Rev. A, {\bf 80}, 043607 (2009).

\bibitem{shore1991}
B. W. Shore, P. Meystre, and S. Stenholm, J. Opt. Soc. Am B {\bf 8}, 903 (1991).

\bibitem{dominic2005}
D. Meiser, C. P. Search, and P. Meystre, Phys. Rev. A, {\bf 71}, 013404 (2005).

\bibitem{mekhov2009}
I. B. Mekhov and H. Ritsch, Laser Phys. {\bf 19}, 610 (2009)

\bibitem{Inouye1999}
S. Inouye {\it et al.}, Science {\bf 285}, 571 (1999).

\bibitem{Schneble2003}
D. Schneble {\it et al.}, Science {\bf 300}, 475 (2003).

\bibitem{Schneble2004}
D. Schneble {\it et al.}, Phys. Rev. A {\bf 69}, 041601(R) (2004).

\bibitem{Yoshikawa2004}
Y. Yoshikawa {\it et al.}, Phys. Rev. A {\bf 69}, 041603(R) (2004).

\bibitem{Moore1999}
M. G. Moore and P. Meystre, Phys. Rev. Lett. {\bf 83}, 5202 (1999); M. G. Moore and P. Meystre, Phys. Rev. A {\bf 58}, 3248 (1998).

\bibitem{Piovella2001}
N. Piovella, M. Gatelli and R. Bonifacio, Opt. Commun. {\bf 194}, 167 (2001).

\bibitem{Robb2005}
G. R. M. Robb, N. Piovella and R. Bonifacio, J. Opt. B {\bf 7}, 93 (2005).

\bibitem{Zobay2005}
O. Zobay and G. M. Nikolopoulos, Phys. Rev. A {\bf 72}, 041604(R) (2005).

\bibitem{Uys2007}
H. Uys and P. Meystre, Phys. Rev. A {\bf 75}, 033805 (2007).

\bibitem{Bonifacio1994}
R. Bonifacio and L. De Salvo, Nucl. Instr. Meth. Phys. Res. Sec. A {\bf 341}, 360 (1994).

\bibitem{Bonifacio21994}
R. Bonifacio {\it et al.}, Phys. Rev. A {\bf 50}, 1716 (1994).

\bibitem{Bonifacio1995}
R. Bonifacio and L. De Salvo, Opt. Commun. {\bf 115}, 505 (1995).

\bibitem{Bonifacio1997}
R. Bonifacio {\it et al.} Phys. Rev. A {\bf 56}, 912 (1997).

\bibitem{Lippi1996}
G. L. Lippi {\it et al.}, Phys. Rev. Lett. {\bf 76}, 2452 (1996).

\bibitem{Hemmer1996}
P. R. Hemmer {\it et al.}, Phys. Rev. lett. {\bf 77}, 1468 (1996).

\bibitem{Kruse2003}
D. Kruse {\it et al.}, Phys. Rev. Lett. {\bf 91}, 183601 (2003).

\bibitem{Cube2004}
C. von Cube {\it et al.}, Phys. Rev. lett. {\bf 93}, 083601 (2004).

\bibitem{Uys2008}
H. Uys and P. Meystre, Laser Phys. Lett. {\bf 5}, 1 (2008)

\bibitem{Elsasser2004}
Th. Els{\"a}sser, B. Nagorny and A. Hemmerich, Phys. Rev. A {\bf 69}, 033403 (2004).

\bibitem{kaplan1985}
A. E. Kaplan and C. T. Law, IEEE J. Quantum Electron., {\bf 21}, 1529 (1985).

\bibitem{pierre1986}
F. Marquis, P. Dobiasch, P. Meystre, and E. M. Wright, J. Opt. Soc. Am. B., {\bf 3}, 50 (1986).

\bibitem{Szirmai2009}
G. Szirmai, D. Nagy, and P. Domokos, Phys. Rev. Lett. {\bf 102}, 080401 (1999).

\bibitem{Szirmai2010}
G. Szirmai, D. Nagy, and P. Domokos, arXiv:1001.1818 (2010).

\bibitem{kauffmann1987}
E. X. DeJesus and C. Kauffmann, Phys. Rev. A, {\bf 35}, 5288 (1987).

\bibitem{kaplan1982}
A. E. Kaplan and P. Meystre, Opt. Comm., {\bf 40}, 229 (1982).



\end{thebibliography}
\end{document}